\RequirePackage{ifpdf}
\ifpdf 
\documentclass[pdftex]{sigma}
\else
\documentclass{sigma}
\fi

\font\frak=eufm10 scaled\magstep1 
 
 2 
scaled\magstep3  \font\tenfrak=eufm10

\def\goth #1{\hbox{{\frak#1}}}

\font\tenfrak=eufm10  \font\sevenfrak=eufm7  \font\fivefrak=eufm5
\newfam\frakfam
\textfont\frakfam=\tenfrak \scriptfont\frakfam=\sevenfrak
\scriptscriptfont\frakfam=\fivefrak
\def\frak{\fam\frakfam\tenfrak}

\def\Im{\mathop{\rm Im}\nolimits}

\def\al{\alpha}
\def\la{\lambda}
\def\om{\omega}
\def\k{\kappa}        

\def\wh{\widehat}
\def\dfrac#1#2{{\displaystyle\frac{#1}{#2}}}

\def\pd#1#2{\frac{\partial #1}{\partial #2}}

\def\Rea{{\rm Re }}
\def\Cos{\mathop{\rm C}\nolimits}    
\def\Sin{\mathop{\rm S}\nolimits}    
\def\Tan{\mathop{\rm T}\nolimits}    
\def\ket#1{|#1\rangle}

\begin{document}

\allowdisplaybreaks

\renewcommand{\PaperNumber}{030}

\renewcommand{\thefootnote}{$\star$}

\FirstPageHeading

\ShortArticleName{A Super-Integrable Two-Dimensional Non-Linear
Oscillator}

\ArticleName{A Super-Integrable Two-Dimensional Non-Linear\\
Oscillator with an Exactly Solvable Quantum Analog\footnote{This
paper is a contribution to the Proceedings of the Workshop on
Geometric Aspects of Integ\-rable Systems
 (July 17--19, 2006, University of Coimbra, Portugal).
The full collection is available at
\href{http://www.emis.de/journals/SIGMA/Coimbra2006.html}{http://www.emis.de/journals/SIGMA/Coimbra2006.html}}}

\Author{Jos\'e F. CARI\~NENA~$^\dag$, Manuel F. RA\~NADA~$^\dag$
and Mariano SANTANDER~$^\ddag$}

\AuthorNameForHeading{J.F. Cari\~nena, M.F. Ra\~nada and M.
Santander}

\Address{$^\dag$~{Departamento de F\'{\i}sica Te\'orica,
Facultad de Ciencias} \\
\phantom{$^\dag$}~{Universidad de Zaragoza, 50009 Zaragoza,
Spain}}

\EmailD{\href{mailto:jfc@unizar.es}{jfc@unizar.es},
\href{mailto:mfran@unizar.es}{mfran@unizar.es}}

\Address{$^\ddag$~{Departamento de F\'{\i}sica Te\'orica,
Facultad de Ciencias} \\
\phantom{$^\ddag$}~{Universidad de Valladolid,  47011 Valladolid,
Spain}} \EmailD{\href{mailto:msn@fta.uva.es}{msn@fta.uva.es}}

\ArticleDates{Received October 31, 2006, in f\/inal form January
24, 2007; Published online February 24, 2007}

\Abstract{Two super-integrable and super-separable classical
systems which can be considered as deformations of the harmonic
oscillator and the Smorodinsky--Winternitz in two dimensions are
studied and identif\/ied with motions in spaces of constant
curvature, the deformation parameter being related with the
curvature. In this sense these systems are to be considered as a
harmonic oscillator and a Smorodinsky--Winternitz system in such
bi-dimensional spaces of constant curvature. The quantization of
the f\/irst system will be carried out and it is shown that it is
super-solvable in the sense that the Schr\"odinger equation
reduces,  in three dif\/ferent coordinate systems, to two separate
equations involving only  one degree of freedom.}


\Classification{37J35; 34A34; 34C15; 70H06}

\section{Super-integrable systems}

There are few integrable systems in the Arnold--Liouville sense:
Hamiltonian systems in a $2n$-di\-mensional symplectic manifold
for which there exist $n$ functionally independent constants of
motion $f_i$  in involution (including the Hamiltonian $H$
itself), i.e.
\begin{gather*}
\{f_i,f_j\}=0,\qquad \forall \; i,j=1,\ldots,n, \qquad
df_1\wedge\cdots\wedge df_n\ne 0.
\end{gather*}
The system is said to be super-integrable when it is integrable
and there exists a set of $m>n$ functionally independent constants
of motion, i.e.\ it possesses more independent f\/irst integrals
than degrees of freedom. The existence of  these additional
f\/irst integrals gives rise to a higher degree of regularity in
the phase space (e.g.\ there exist periodic orbits) since the
trajectories are restricted to submanifolds of dimension lower
than $n$. In particular,  a system with $n$ degrees of freedom
possessing $2n-1$ independent f\/irst integrals is said to be
maximally super-integrable.

It is also well-known  \cite{Be1873} that the only central
potentials in which all bounded orbits are closed (periodic) are
given by $V = (1/2) \omega_0^2 r^2$ and $V=-k/r$.  From a modern
point of view the existence of closed trajectories is considered
as a consequence of the existence of the  maximal number of
functionally independent integrals of motion; thus, the result
obtained by Bertrand is a proof of the super-integrability of
these two systems.  In fact, the harmonic oscillator shares with
the Kepler problem a very distinguished property not only in
Classical but also in Quantum Mechanics, where super-integrability
leads to energy levels depending on a single quantum number.

   Fris {\it et al.} studied in 1965 the Euclidean $n=2$ systems
which admit separability in two dif\/ferent coordinate systems
\cite{FrMS65}, and obtained four families $V_r$, $r=a,b,c,d$, of
super-integrable potentials with constants of motion linear or
quadratic in the velocities (momenta). The two f\/irst families
\begin{gather*}
   V_a = \frac{1}{2}\om_0^2(x^2 + y^2) + \frac{k_2}{x^2}+ \frac{k_3}{y^2},\qquad
   V_b = \frac{1}{2}\om_0^2(4 x^2 + y^2) + k_2 x + \frac{k_3}{y^2},
\end{gather*}
can be considered as the more general Euclidean deformations (with
strengths $k_2$, $k_3$) of the $1:1$ and $2:1$ harmonic
oscillators preserving quadratic super-integrability (the other
two families, $V_c$ and $V_d$, were related with the Kepler
problem). The super-integrability of $V_a$, which is known as the
`Smorodinsky--Winternitz' (S-W) potential, has been studied by
Evans \cite{Ev90PhLet,Ev91Jmp} for the general case of $n$ degrees
of freedom.

A large number of papers have been published on
super-integrability in these last years, most of them related with
quadratic superintegrabilty  (see~\cite{Montr04} for the
proceedings of a workshop on super-integrability and Refs.
\cite{KaKrMi05a,KaKrMi05b,BlSe05,DaskYp06} for some very recent
studies on super-integrable systems with integrals quadratic in
momenta). The idea is that if we call super-separable a~system
that admits Hamilton--Jacobi separation of variables
(Schr\"odinger in the quantum case) in more than one coordinate
system, then quadratic super-integrability (i.e.,
super-integrability with linear or quadratic constants of motion)
can be considered as a property arising from super-separability.
We note that these studies also include non-Euclidean Hamiltonian
systems
\cite{GrPoSi95,Ra97,KaMiPo97,RaSa99,KaMiPo00,Sl00,KaKrPo01,RaSa02I,KaKrWi02,BaHeSa03,RaSa03II,BaHeRa03,HeBa06Sigma}
and that in both cases, Euclidean and non-Euclidean, many of these
systems are closely related with the harmonic oscillator.

The rareness of integrable systems leads to the following
question.
  Is it possible to deform a super-integrable system but
preserve  super-integrability of the system? We report here some
results of previous works where such question is af\/f\/irmatively
answered. We shall show that this is possible for two important
examples: the isotropic bi-dimensional harmonic oscillator and the
Smorodinsky--Winternitz system.

  A very interesting example of a one-dimensional  nonlinear
oscillator depending of a parameter $\lambda$ was studied in 1974
by Mathews and Lakshmanan \cite{MaLa74,LaRa03} and  it has
recently been proved~\cite{CaRaSS04} that this particular
nonlinear system can be generalized to the two-dimensional case,
and even to the $n$-dimensional case:  these higher dimensional
systems are super-integrable deformations admitting  $2n-1$
quadratic constants of motion. It is also proved that there is  a
related $\la$-de\-pendent two-dimensional oscillator isotonic
oscillator that  is a super-integrable deformation, i.e.\ a
  $\la$-dependent version of the
Smorodinski--Winternitz system \cite{CaRaS05,CaRaSDubna06}.
Actually,
  the deformation introduced by the parameter $\la$
modif\/ies the Hamilton--Jacobi equation but preserves the
existence of a multiple separability. Moreover, we point out that
a geometric interpretation of the higher-dimensional systems was
proposed in relation with the dynamics on spaces of constant
curvature.

   This paper must be considered as a survey summarizing recent works
by the authors on properties related with the integrability and
super-integrability of certain two-dimensional $\la$-de\-pen\-dent
systems related with the harmonic oscillator and it is neither a
review of the whole f\/ield nor a comparison with other approaches
of dif\/ferent authors.  It is mainly focused on the study of
deformations of super-integrable systems that do not alter the
super-integrability structure.  It starts with a brief review of
the basic properties of the above mentioned isotonic oscillator,
S-W system and Mathews--Lakshmanan oscillator and some
two-dimensional $\la$-dependent oscillator-like systems, f\/irst
in in  the classical approach and afterwards in their quantum
counterparts.

   Roughly speaking this survey is divided into three main parts: f\/irst,
the analysis of the classical $\la$-dependent oscillator including
the existence of Hamilton--Jacobi multiple separability (existence
of alternative coordinate systems in which the corresponding
$\la$-dependent Hamilton--Jacobi equation separates). Second,  the
relation of this $\la$-dependent nonlinear model with the harmonic
oscillator on the three spaces of constant curvature $(S^2,{\Bbb
E}^2,H^2)$. The f\/inal part is devoted to  the analysis of these
systems from the quantum viewpoint.

   The second part presents a geometrical approach and proves
that these systems can be considered in two dif\/ferent ways:
either as a nonlinear deformation of a linear system, or simply as
a model of the oscillator on spaces of constant curvature. In the
f\/irst case the parameter $\la$ represents the strength of the
deformation and in the second one the curvature of the space.
Finally, in the third part,  devoted to the quantum version of
this $\la$-dependent oscillator, it is proved that the system is
exactly solvable:
  the Hamiltonian can be factorized and  the wave functions
and energies can be explicitly obtained.  It includes some points
such as: (i) Analysis of the transition from the classical
$\la$-dependent system to the quantum one, (ii) Exact resolution
of the $\la$-dependent Schr\"odinger equation, factorization
method and existence of operators $A$ and $A^+$, and
shape-invariance property, (iii) Schr\"odinger multiple
separability and quantum super-integrability.

In more detail the structure of the paper is as follows: In next
section we describe shortly two one-dimensional classical systems
which can be considered as generalizations of the harmonic
oscillator and the isotonic oscillator: the $\lambda$-deformed
nonlinear oscillator and the deformed isotonic oscillator. Section
3 deals with two-dimensional generalizations of these classical
systems and their deformations and Section 4 studies the
separability properties of their Hamilton Jacobi equations. A
geometric interpretation is given in Section 5 and a quantization
of the nonlinear oscillator, is carried out in Section 6, and the
quantum spectrum s computed by using the traditional power series
expansion. The system is shown in Section 7 to admit a shape
invariant factorization which allows us to explicitly compute the
spectrum in an algebraic alternative way. Finally, in Section 8 we
sketch the method to be used for the corresponding two-dimensional
system and prove that he Hamiltonian can be written as a sum of
three dif\/ferent terms such that each one commutes with the sum
of the other two, what provides us alternative complete sets of
compatible observables.

\section{Some one-dimensional classical systems}

We start this section by reviewing the basic properties of two
simple classical systems.

\subsection{The harmonic oscillator}

The dynamics of the classical harmonic oscillator in one dimension
is given by
\begin{gather*}
\dfrac {dx}{dt}=v, \qquad \dfrac {dv}{dt}=-\omega^2 x
\end{gather*}
and is described by a Lagrangian $L=(1/2) \left(v^2-\omega^2
x^2\right)$. A complex variable $z = \omega x + i v$ may be
introduced and  the equations of the motion become
\begin{gather*}
\frac {dz}{dt}=-i\omega z,
\end{gather*}
whose general solution is $z = z_0 e^{-i\omega t}$, i.e.
\begin{gather*}
x=x_0  \cos \omega t-\frac{v_0}\omega \sin\omega t = A\cos(\omega
t+\varphi),
\end{gather*}
and therefore the solutions are periodic with angular frequency
$\omega$, while $A$ and $\varphi$ are arbit\-rary.

\subsection{The isotonic harmonic oscillator}

The isotonic oscillator is described by the Lagrangian
\cite{Cal69,Pe90}
\begin{gather*}
   L  = \frac{1}{2} v_x^2 -\frac{1}{2} {\alpha}^2 x^2 - \frac{k}{x^2},
   \qquad  k>0,
\end{gather*}
(like a harmonic oscillator plus a centripetal barrier). This is
an important example of an isochronous system and actually they
are the only two rational potentials giving rise to
iso\-chro\-nous systems \cite{ChaVes05,{ACMP07}}.

The Euler--Lagrange  equation
\begin{gather*}  \ddot{x}  + {\alpha}^2 x + \frac{c}{x^3} = 0
,{\qquad} c = -2  k,
\end{gather*}
is a particular case of the so-called Pinney--Ermakov equation
\cite{Pin50,{ermakov}} whose general solution can be written as $x
= ({1}/(\alpha A) \sqrt{ (\alpha^2A^4 + c)\sin^2({\alpha}t+\phi)
-c}$.

The corresponding quantum system  admits a shape-invariant
factorization \cite{sifisot} and therefore is solvable by means of
algebraic methods.

\subsection{A 1-dimensional  nonlinear oscillator}

   In 1974 Mathews and Lakshmanan \cite{MaLa74} studied the equation of
motion
\begin{gather}
    \big(1 +\lambda x^2\big) \ddot{x}-\lambda x \dot{x}^2 + \alpha^2 x  = 0
,\qquad\lambda>0.\label{nloeq}
\end{gather}
The general solution takes the form $x  = A \sin(\omega t +
\phi)$, with the following additional restriction linking angular
frequency $\omega$ and amplitude $A$:
\begin{gather*}
   \omega^2  = \frac{\alpha^2}{1 + \lambda A^2} .
\end{gather*}
The equation (\ref{nloeq}) is the Euler--Lagrange equation for the
Lagrangian:
\begin{gather*}
   L_\lambda(x,\dot x)  =
   \frac{1}{2} \frac{1}{1 + \lambda x^2}  \big(\dot{x}^2 - \alpha^2 x^2\big).
\end{gather*}
It describes a system with nonlinear oscillations with an
amplitude dependent frequency (or period). We can also allow
negative values for $\lambda$ \cite{CaRaSS04}, but when
$\lambda<0$  the values of $x$ are limited by the condition
$|x|<1/\sqrt{|\lambda|}$. In the limit $\lambda\to 0$ we recover
the equation of motion and both the Lagrangian of the harmonic
oscillator and the frequency become independent of the amplitude.
Therefore, the system can be seen as a deformation of the harmonic
oscillator.

It  can also  be seen as an oscillator with a position-dependent
ef\/fective mass which depends on $\lambda$:
\begin{gather*}m_\lambda= \frac{1}{1 + \lambda x^2}.
\end{gather*}

The Hamiltonian for such a system is:
\begin{gather*}
    H_\lambda(x,p)  = \frac{1}{2}\big(1 + \lambda x^2\big)  p^2 + \frac{1}{2}
\frac{\alpha^2 x^2}{1 + \lambda x^2}.
\end{gather*}
The important fact is that there is an interesting generalization
to $n=2$ or even arbitrary $n$~\cite{CaRaSS04}.

Note that the Lagrangian for the  one-dimensional free-particle
(i.e. for $\alpha=0$)
\begin{gather*}
   L_\lambda(x,v_x) = T_1(\lambda) =
   \frac{1}{2} \frac{v_x^2}{1 + \lambda x^2}
\end{gather*}
is invariant under the  vector f\/ield
\begin{gather*}
   X_x^t(\lambda) = \sqrt{ 1+\lambda x^2 }  \pd{}{x}
   + \frac{\lambda x v_x}{\sqrt{1+\lambda x^2 }} \pd{}{v_x} .
\end{gather*}
which is the natural lift to the phase space
${\mathbb{R}}\times{\mathbb{R}}$ of the vector f\/ield in
$\mathbb{R}$
\begin{gather}
    X_x(\lambda) = \sqrt{ 1+\lambda x^2 }  \pd{}{x},
    \label{Xx1}
\end{gather}
i.e.\ $X_x^t(\lambda)\bigl(T_1(\lambda)\bigr)=0$.

\subsection{A deformed isotonic oscillator}

We study next a deformed isotonic oscillator in one dimension
described by the Lagrangian \cite{CaRaS05}:
\begin{gather*}
   L_\la(x,v_x,k) = \frac{1}{2} \frac{v_x^2}{1 + \la x^2}
   - \frac{1}{2} \frac{\alpha^2 x^2}{1 + \lambda x^2}
   - \frac{k}{x^2}.
\end{gather*}
Here $\lambda$ can be any real number, but when $\lambda<0$ the
possible values of $x$ are such that $|x|<1/\sqrt{|\lambda|}$.

The Euler--Lagrange equation is
\begin{gather*}
   \frac{d^2x}{dt^2} - \frac{\la x}{1 + \la x^2} \left(\frac{d
   x}{dt}\right)^2 +  \frac{\alpha^2 x}{1 + \la x^2}
   - 2 k  \frac{1 + \la x^2}{x^3} = 0,
\end{gather*}
and one can see that the general solution for bounded motions is
\cite{CaRaS05}:
\begin{gather*}
x = \frac{1}{\om A} \sqrt{(\om^2A^4 - 2k) \sin^2(\om t + \phi) +
2k},
\end{gather*}
where
\begin{gather*}
\om R_1 \bigl[ 2k +  (\om^2A^4 - 2k) \sin^2(\om t + \phi) \bigr]^2
= 0
 \end{gather*}
 with
\begin{gather*}
   R_1 = \la \om^2 A^4 - \big(\alpha^2 - \om^2 -  2 k \la^2\big) A^2
   + 2 k \la .
\end{gather*}

It can be done similarly for unbounded motions, the functions
$\sin(\omega  t+\phi)$ being then replaced by functions
$\sinh(\Omega  t+\phi)$. There also exist limit unbounded motions
of the form
\begin{gather*}  x = \sqrt{(A t + B)^2 + C } .
\end{gather*}

It will be shown that as  in the harmonic oscillator case, there
is an interesting generalization to the $n=2$ case.

\section{Some two-dimensional classical systems}
\subsection{The harmonic oscillator}

The Hamiltonian of a  two-dimensional classical harmonic
oscillator is (for simplicity $m=1$):
\begin{gather*}
   H(x,y,p_x,p_y) = \frac{1}{2} \big(p_x^2 + p_y^2\big)
   + \frac{1}{2} \big(\omega_1^2  x^2 + \omega_2^2  y^2\big),
\end{gather*}
and we can easily check that the energy functions for each degree
of freedom are constants of motion:
\begin{gather*} I_1=E_x=\frac 12 \big(p_x^2+\omega_1^2  x^2\big), \qquad
   I_2=E_y=\frac 12 \big(p_v^2+\omega_2^2  y^2\big).
\end{gather*}
The rational case, for which $\omega_1 = n_1 {\omega_0}$,
$\omega_2 = n_2 {\omega_0}$, with $n_1,n_2 \in \mathbb{N}$, is
super-integrable \cite{CMR02}. In fact, let $K_x$  and  $K_y$ be
def\/ined by $K_x = p_x + i  n_1 {\omega_0}  x$, $K_y = p_y + i
n_2 {\omega_0} y$.  The Hamiltonian $H$ and the canonical
symplectic form $\Omega_0$ can be expressed in terms of such
functions as follows:
\begin{gather*}
   H = \frac 12\left(K_x  K_x^*+K_y  K_y^*\right),\qquad
   \Omega_0 = \frac{i }{2 n_1 \omega_0} {dK_x\wedge dK_x^*}
   + \frac{i }{2 n_2 \omega_0} {dK_y\wedge dK^*_y},
\end{gather*}
and  therefore, as the fundamental Poisson brackets are
$\{K_x,K^*_x\} = 2 i   n_1 \omega_0$ and $\{K_y,K^*_y\} = 2 i  n_2
\omega_0$, the evolution equations are
\begin{gather*}
   \frac{d}{d t} K_x   =   i  n_1  {\omega_0} K_x,{\qquad}
   \frac{d}{d t} K_y^* = - i  n_2  {\omega_0} K_y^*.
   \end{gather*}

Hence, the complex function $J$ def\/ined as $J_{n_1,n_2} =
K_x^{n_2} (K_y^{*})^{n_1}$ is a (complex) constant of motion which
determines two dif\/ferent real f\/irst integrals: $I_3 = \Im
(J_{n_1,n_2})$ and $I_4 = \Rea (J_{n_1,n_2})$. They are
polynomials in the momenta of degrees $n_1+n_2-1$ and $n_1+n_2$,
respectively.

Two particularly important examples are the isotropic harmonic
oscillator (for $n_1=n_2=1$) and the case $n_1=1$, $n_2=2$. In
these cases the four constants are not independent but only three
of them are independent. For instance, in the isotropic case the
constants of motion take the form $I_3=p_xp_y+\omega^2xy$ and $
I_4= xp_y-yp_x$.

\subsection[The Smorodinsky-Winternitz system]{The Smorodinsky--Winternitz system}

A 2-dimensional generalization of the isotonic oscillator with
rotational symmetry would be
\begin{gather*}
   V_{ri} = \frac{1}{2} \om_0^2\big(x^2 + y^2\big) + \frac{k_1}{x^2+ y^2}.
\end{gather*}
There is however another super-integrable generalization given by
the Smorodinsky--Winternitz potential \cite{FrMS65}
\begin{gather*}
   V_{SW} = \frac{1}{2} \om_0^2\big(x^2 + y^2\big)
          + \frac{k_2}{x^2} + \frac{k_3}{y^2},
\end{gather*}
which is a set of two non-interacting isotonic systems with the
same frequency and in general dif\/ferent constants $k_2\ne k_3$.
Of course, the case $k_2=k_3=0$ reduces to the usual isotropic
harmonic oscillator. The energies of each degree of freedom are
constants of the motion and there exists a third constant of
motion given by
\begin{gather*}
   C = J^2+k_2 \frac{y^2}{x^2}-k_3 \frac{x^2}{y^2},
   \qquad {\rm with }\qquad J = xv_y-yv_x.
\end{gather*}

\subsection{A nonlinear oscillator in two dimensions}

It has recently been proved \cite{CaRaSS04} that there exists a
1-parameter dependent generalization of the nonlinear oscillator
for the 2-dimensional case with the following requirements:

1.~The kinetic term $T_{2}(\lambda)$ is a quadratic function of
the velocities  invariant under rotations.

{\samepage 2.~$T_2(\lambda)$ is invariant under (the tangent lifts
of) the vector f\/ields $X_1(\lambda)$ and $X_2(\lambda)$ given by
\begin{gather*}
     X_1(\lambda) = \sqrt{ 1+\lambda r^2 }  \pd{}{x},\qquad
     X_2(\lambda) = \sqrt{ 1+\lambda r^2 }  \pd{}{y},
\end{gather*}
which are extensions to ${\mathbb{R}}^2$ of the vector f\/ield
$X_x(\lambda)$ in the $n=1$ case given by (\ref{Xx1}).

}

These conditions lead to the following form for the kinetic
energy:
\begin{gather*}
   T_2(\lambda) = \frac{1}{2} \frac{1}{1 + \lambda r^2}
   \bigl[ v_x^2 + v_y^2 + \lambda (x v_y - y v_x)^2  \bigr],\qquad
   r^2 = x^2+y^2.
\end{gather*}
The term $\lambda (x v_y - y v_x)^2$ represents a two-dimensional
contribution that cannot appear in the one-dimensional case. When
$\lambda<0$ this function  will have a singularity at $1 -
|\lambda| r^2=0$ and we should restrict our dynamics to the
interior of the circle $x^2+y^2<1/|\lambda|$ where $T_2(\lambda)$
is positive def\/inite.

The kinetic energy  $T_2(\lambda)$ is determined by the
$\lambda$-dependent metric
\begin{gather*}
   ds^2(\lambda) = \frac{1}{1 + \lambda r^2}
   \bigl[ \big(1 + \lambda y^2\big) dx^2 + \big(1 + \lambda x^2\big) dy^2
   - 2 \lambda x y  dx dy \bigr].
\end{gather*}
$T_2(\lambda)$ remains invariant under the actions of the lifts of
the vector f\/ields $X_1(\lambda)$, $X_2(\lambda)$, and $X_J$,
given by
\begin{gather*}
   X_1(\lambda) = \sqrt{ 1+\lambda r^2 }  \pd{}{x},\qquad
   X_2(\lambda) = \sqrt{ 1+\lambda r^2 }  \pd{}{y},\qquad
   X_J   = x \pd{}{y} - y \pd{}{x},
\end{gather*}
namely
\begin{gather*}
X_1^t(\lambda) = \sqrt{ 1+\lambda r^2 }  \pd{}{x}
   + \lambda \frac{x v_x + y v_y}{\sqrt{1+\lambda r^2 }} \pd{}{v_x},
\qquad X_2^t(\lambda) = \sqrt{ 1+\lambda r^2 }  \pd{}{y}
   + \lambda \frac{x v_x + y v_y}{\sqrt{1+\lambda r^2 }} \pd{}{v_y},
\\
X_J^t   = x \pd{}{y} - y \pd{}{x}+ v_x \pd{}{v_y} -v_y \pd{}{v_x},
\end{gather*}

   These vector f\/ields close on a Lie algebra:
\begin{gather*}
   [X_1(\lambda),X_2(\lambda)] = {\lambda} X_J,{\qquad}
   [X_1(\lambda),X_J] =      X_2(\lambda),{\qquad}
   [X_2(\lambda),X_J]=    - X_1(\lambda),
\end{gather*}
which is either isomorphic to ${\goth{so}}(3,{\mathbb{R}})$ when
$\lambda>0$, to ${\goth{so}}(2,1)$ when $\lambda<0$, or   to the
Lie algebra of the Euclidean group for $\lambda=0$.

The appropriate generalization for the potential of the nonlinear
two-dimensional $\la$-dependent oscillator is given by
\begin{gather*}
   V_{\la}(x,y)  = \frac{\alpha^2}{2} \frac{x^2+y^2}{1 + \la (x^2+y^2)} .
\end{gather*}

   This bi-dimensional nonlinear oscillator is completely integrable
\cite{CaRaSS04}, because one can show that, if $K_1$ and  $K_2$
are the  functions
\begin{gather*}  K_1 = P_1(\lambda) + {i }{\alpha}\
   \frac{x}{\sqrt{ 1 + \lambda r^2 }},\qquad
   K_2 = P_2(\lambda) + {i }{\alpha}\
   \frac{y}{\sqrt{ 1 + \lambda r^2 }},
\end{gather*}
with
\begin{gather*}
    P_1(\lambda)
    = \frac{v_x - \lambda J y}{\sqrt{ 1 + \lambda r^2 }},\qquad
    P_2(\lambda)
    = \frac{v_y + \lambda J x}{\sqrt{ 1 + \lambda r^2 }}, \qquad
    J=x v_y - y v_x,
\end{gather*}
then the complex functions  $K_{ij}$ def\/ined as $K_{ij} = K_i
K_j^*$,  $ i,j=1,2$, are constants of motion.

   In fact, the time-evolution of the functions $K_1$ and $K_2$ is
\begin{gather*}   \frac{d}{d t} K_1
   =  \frac{{i}{\alpha}}{1 + \lambda r^2}  K_1,\qquad
   \frac{d}{d t} K_2  =
   \frac{{i}{\alpha}}{1 + \lambda r^2} K_2,
\end{gather*}
from which we see that the complex functions $K_{ij}$ are
constants of the motion.

Therefore the system is super-integrable with the following
f\/irst integrals of motion:
\begin{gather*}
   I_1(\lambda) = | K_1 |^2,{\qquad}
   I_2(\lambda) = | K_2 |^2,{\qquad}
   I_3 = \Im(K_{12}) = {\alpha} (x v_y - y v_x)  .
\end{gather*}

\subsection[Nonlinear Smorodinsky-Winternitz system]{Nonlinear Smorodinsky--Winternitz system}

We can consider a deformed Smorodinsky--Winternitz system with
Lagrangian \cite{CaRaS05}
\begin{gather*}
   L = \frac{1}{2(1 + \lambda r^2)}
   \bigl[v_x^2 + v_y^2 + \lambda (x v_y - y v_x)^2 \bigr]
   - \frac{\alpha^2}{2} \frac{x^2+y^2}{1 + \la (x^2+y^2)}
   + \frac{k_2}{x^2} + \frac{k_3}{y^2} .
\end{gather*}
The E-L equations cannot be directly solved in a simple way, but
one can check by direct computation the existence of three
independent constants of motion:
\begin{gather*}
I_1 = P_1^2 + \frac{\alpha^2 x^2}{1 + \la (x^2+y^2)}
    + 2 k_2 \frac{1 + \la y^2}{x^2},
\\
I_2 = P_2^2 + \frac{\alpha^2 y^2}{1 + \la (x^2+y^2)}
    + 2 k_3 \frac{1 + \la x^2}{y^2},
\qquad I_3 = J^2 + 2 k_2  \frac{y^2}{x^2} + 2 k_3
\frac{x^2}{y^2}.
\end{gather*}

This proves that for any value of $\lambda$ this system is
completely integrable. Note that  for $\la=0$ the previous
constants of motion  reduce to the three f\/irst integrals of the
S-W potential. Consequently, this system is a  deformation of the
S-W potential which preserves the completely integrability.

\section[Hamilton-Jacobi super-separability of the previous systems]{Hamilton--Jacobi super-separability of the previous systems}

The Legendre transformation for a Lagrangian $L_\lambda =
T_2(\lambda) - V(x)$ is given by \cite{CaRaSS04}
\begin{gather*}
   p_x = \frac{(1 + \lambda y^2) v_x - \lambda x y v_y}{1 +
\lambda r^2},\qquad
   p_y = \frac{(1 + \lambda x^2) v_y - \lambda x y v_x}{1 + \lambda r^2} .
\end{gather*}
Note that  $x p_y - y p_x=x v_y - y v_x$.

The general expression for a $\lambda$-dependent Hamiltonian is
\begin{gather*}
   H(\lambda) = \frac{1}{2} \bigl[ p_x^2 + p_y^2
   + \lambda (x p_x + y p_y)^2 \bigr] + \frac{1}{2} {\alpha^2} V(x,y),
\end{gather*}
and hence the associated Hamilton--Jacobi equation is
\begin{gather*}
   \left(\pd{S}{x}\right)^2 + \left(\pd{S}{y}\right)^2
   + \lambda \left(x \pd{S}{x} + y \pd{S}{y}\right)^2
   + \alpha^2 V(x,y) = 2 E .
\end{gather*}
This equation is not separable in $(x,y)$ coordinates but there
exist three particular orthogonal coordinate systems, and three
particular families of associated potentials, for which such
a~Hamiltonian  admits Hamilton--Jacobi separability:

(i) In terms of the new coordinates $(z_x,y)$, $z_x = {x}/{\sqrt{
1 + \lambda y^2 }}$, the Hamilton--Jacobi equation becomes:
\begin{gather*}
   \big(1 + \lambda z_x^2\big)\left(\pd{S}{z_x}\right)^2
   + \big(1 + \lambda y^2\big)^2\left(\pd{S}{y}\right)^2 +\alpha^2 \big(1 + \lambda y^2\big)V
   = 2 \big(1 + \lambda y^2\big) E
\end{gather*}
so if the potential $V(x,y)$ can be written in the form
\begin{gather*}
   V = \frac{W_1(z_x)}{ 1 + \lambda y^2} + W_2(y)
\end{gather*}
then the equation becomes separable.

The potential is therefore integrable with the following two
quadratic integrals of motion
\begin{gather*}
I_1(\lambda) = \big(1 + \lambda r^2\big)p_x^2 + {\alpha^2}
W_1(z_x),
\\
I_2(\lambda) = \big(1 + \lambda r^2\big)p_y^2 - \lambda J^2 +
{\alpha^2} \biggr(W_2(y) - \frac{\lambda y^2}{ 1 +\lambda y^2}
W_1(z_x)\biggl).
\end{gather*}
and $H_\lambda $ can be written as:
$H_\lambda=(1/2)(I_1(\lambda)+I_2(\lambda))$.

(ii) Similarly, using coordinates $(x,z_y)$, $z_y = {y}/{\sqrt{ 1
+ \lambda x^2 }}$, the Hamilton--Jacobi equation becomes:
\begin{gather*}
   \big(1 + \lambda x^2\big)^2\left(\pd{S}{x}\right)^2
   + \big(1 + \lambda z_y^2\big)\left(\pd{S}{z_y}\right)^2 + \alpha^2 \big(1 +
\lambda x^2\big)V
   = 2 \big(1 + \lambda x^2\big) E,\nonumber
\end{gather*}
therefore, if the potential $V(x,y)$ can be written on the form
\begin{gather*}
   V = W_1(x) + \frac{W_2(z_y)}{ 1 + \lambda x^2}
\end{gather*}
then the equation becomes separable.

The potential is therefore integrable with the following two
quadratic integrals of motion
\begin{gather*}
I_1(\lambda) = \big(1 + \lambda r^2\big)p_x^2 - \lambda J^2
 + {\alpha^2} \left(W_1(x) - \frac{\lambda x^2}{ 1 +\lambda x^2} W_1(z_y)\right),
\\
 I_2(\lambda) = \big(1 + \lambda r^2\big)p_y^2 + {\alpha^2} W_2(z_y)
\end{gather*}
and $H_\lambda $ can be written as: $H_\lambda =
(1/2)(I_1(\lambda) + I_2(\lambda))$.

(iii) In polar coordinates $(r,\phi)$ the Hamiltonian $H(\lambda)$
is
\begin{gather*}
   H(\lambda) = \frac{1}{2} \left[ \big(1 + \lambda r^2\big)p_r^2
   + \frac{p_\phi^2}{r^2} \right]
   + \frac{\alpha^2}{2} V(r,\phi)
\end{gather*}
so that the Hamilton--Jacobi equation is given by
\begin{gather*}
   \big(1 + \lambda r^2\big)\biggl(\pd{S}{r}\biggr)^2
    + \frac{1}{r^2} \biggl(\pd{S}{\phi}\biggr)^2 + \alpha^2 V(r,\phi) = 2 E  .
\end{gather*}
If the potential $V$ is of the form
\begin{gather*}
   V = F(r) + \frac{G(\phi)}{r^2},
\end{gather*}
then the equation admits separability
\begin{gather*}
   r^2\big(1 + \lambda r^2\big)\biggl(\pd{S}{r}\biggr)^2
   + r^2 \bigl( \alpha^2 F(r) - 2 E\bigr)
   + \biggl(\pd{S}{\phi}\biggr)^2  + \alpha^2 G(\phi)  = 0 .
\end{gather*}
The potential $V$ is integrable with the following two quadratic
integrals of motion:
\begin{gather*}
I_1(\lambda) = \big(1 + \lambda r^2\big)p_r^2 + \frac{1-r^2}{r^2}
p_\phi^2 + \alpha^2 \biggl[  F(r) + \frac{1-r^2}{r^2}
G(\phi)\biggr],
\\
I_2(\lambda) = p_\phi^2 + \alpha^2 G(\phi)
\end{gather*}
and $H_\lambda $ can be written as:
$H_\lambda=(1/2)(I_1(\lambda)+I_2(\lambda))$.

\subsection{Super-separability of the nonlinear oscillator}

Using the previous results one can see that the potential
\begin{gather*}
   V_\lambda = \frac{\alpha^2}{2}  \frac{x^2+y^2}{1 + \lambda (x^2+y^2)}
\end{gather*}
is actually super-separable (separable in more than one system of
coordinates) \cite{CaRaSS04}.  In fact, the potential for the
deformed oscillator we are considering can be alternatively
written as follows:
\begin{gather*}
V_\lambda  = \frac{\alpha^2}{2} \frac{1}{1+\lambda y^2}
\left[\frac{z_x^2}{1+\lambda z_x^2}  +  y^2 \right]
  =\frac{\alpha^2}{2} \frac{1}{1+\lambda x^2}
   \left[ x^2  +  \frac{z_y^2}{1+\lambda z_y^2} \right]
=\frac{\alpha^2}{2} \frac{r^2}{1+\lambda r^2}.
\end{gather*}

Consequently, the Hamiltonian
\begin{gather*}
H(\lambda) = \frac{1}{2} \bigl[ p_x^2 + p_y^2 + \lambda (x p_x+ y
p_y)^2 \bigr] + \frac{\alpha^2}{2}  \frac{x^2+y^2}{1 + \lambda
(x^2+y^2)}
\end{gather*}
admits the following decomposition
\begin{gather*}
H(\lambda) = H_1(\lambda) + H_2(\lambda) - \lambda H_3
\end{gather*}
where the three partial functions $H_1$, $H_2$, and $H_3$ are:
\begin{gather*}
H_1(\lambda)  = \frac{1}{2} \left[\big(1 + \lambda r^2\big) p_x^2
+ \alpha^2\frac{x^2}{1+\lambda r^2}\right],
\\
H_2(\lambda)  =  \frac{1}{2} \left[\big(1 + \lambda r^2\big) p_y^2
+ \alpha^2\frac{y^2}{1+\lambda r^2}\right], \qquad H_3  =
\frac{1}{2} (x p_y - y p_x)^2,
\end{gather*}
and each one  has a vanishing Poisson bracket with $H$,
\begin{gather*}
   \bigl\{H,H_1(\lambda)\bigr\} = 0,\qquad
   \bigl\{H,H_2(\lambda)\bigr\} = 0,\qquad
   \bigl\{H,H_3(\lambda)\bigr\} = 0.
\end{gather*}

\subsection{Super-separability of the deformed S-W system}

     The important point is that, in a similar manner, the $\la$-deformed
Smorodinsky--Winternitz potential
\begin{gather*}
   V_{\la,k}  = \frac{\alpha^2}{2} \left(\frac{x^2+y^2}{1 +
\la (x^2+y^2)} \right)
   + \frac{k_2}{x^2} + \frac{k_3}{y^2}
\end{gather*}
can be alternatively written in the following three dif\/ferent
ways \cite{CaRaS05}
\begin{gather*}
     V_{\la,k}  = \frac{\alpha^2}{2} \left(\frac{1}{1+\la y^2}\right)
     \left[\frac{z_x^2}{1+\la z_x^2}  +  y^2 \right]
     + \left(\frac{k_2}{1+\la y^2}\right)\frac{1}{z_x^2} + \frac{k_3}{y^2}
     \\
     \phantom{V_{\la,k}}{}
     =  \frac{\alpha^2}{2} \left(\frac{1}{1+\la x^2}\right)
     \left[ x^2  +  \frac{z_y^2}{1+\la z_y^2} \right]
     + \frac{k_2}{x^2} + \left(\frac{k_3}{1+\la x^2}\right)\frac{1}{z_y^2}
     \\
     \phantom{V_{\la,k}}{}
     =  \frac{\alpha^2}{2} \left(\frac{r^2}{1+\la r^2}\right)
     + \frac{k_2}{r^2\cos^2\phi} + \frac{k_3}{r^2\sin^2\phi}  .
\end{gather*}
Therefore, it is super-separable since it is separable in three
dif\/ferent systems of coordinates, $(z_x,y)$, $(x,z_y)$, and
$(r,\phi)$.  This remarkable property means that the Hamiltonian
\begin{gather*}
H_{\la,k} = \frac{1}{2} \bigl[ p_x^2 + p_y^2   + \la (x p_x + y
p_y)^2 \bigr]  + V_{\la,k}
\end{gather*}
admits the following decomposition $H_{\la,k} = H_{p_x} + H_{p_y}
- \la H_J$, where the three partial functions~$H_1$, $H_2$, and
$H_3$, are given by
\begin{gather*}
     H_{p_x}  = \frac{1}{2} \left[\big(1 + \la r^2\big) p_x^2 + \alpha^2
     \left(\frac{x^2}{1+\la r^2}\right)  \right]
     + k_2\left(\frac{1 + \la y^2}{x^2}\right),
     \\
     H_{p_y}  =  \frac{1}{2} \left[\big(1 + \la r^2\big) p_y^2 + \alpha^2
     \left(\frac{y^2}{1+\la r^2}\right)  \right]
     + k_3\left(\frac{1 + \la x^2}{y^2}\right),
     \\
     H_J =  \frac{1}{2} \bigl(x p_y - y p_x\bigr)^2
     +  k_2  \frac{y^2}{x^2} +  k_3  \frac{x^2}{y^2} .
\end{gather*}

Each one of these three terms has a vanishing Poisson bracket with
$H$ for any value of the parameters $\la$, $k_2$ and $k_3$
\begin{gather*}
    \bigl\{H_{\la,k},H_{p_x} \bigr\} = 0,\qquad
    \bigl\{H_{\la,k},H_{p_y} \bigr\} = 0,\qquad
    \bigl\{H_{\la,k},H_J \bigr\} = 0.
\end{gather*}

Consequently,  the Hamiltonian can be written as a sum, not of
two, but of three integrals of motion. The third one represents
the contribution of the angular momentum $J$ to $H$, with the
parameter $\la$ as a coef\/f\/icient, therefore vanishing  in the
limit $\la\to 0$.

\section{A geometric interpretation}

The existence of additional constants of motion for the harmonic
oscillator in a spherical geometry was studied by Higgs in 1979
\cite{Hi79}. The Higgs approach considers the motion on $S^n$,
embedded in the Euclidean space ${\Bbb E}^{n+1}$, by means of a
central (also known as gnomonic) projection on a plane $\Pi^n$
tangent to $S^n$ at a chosen point. In this way he could study the
properties of the spherical version of the Fradkin tensor. A
dif\/ferent alternative approach is discussed
in~\cite{RaSa02I,RaSa03II} by using curvature-dependent
trigonometric and hyperbolic functions, where use is made of the
curvature $\k$ as a parameter so that the dynamics can be studied
at the same time in the sphere $S^2$ and in the hyperbolic plane
$H^2$.

   In dif\/ferential geometric terms, the three spaces with constant
curvature, sphere $S^2$, Euclidean plane ${\Bbb E}^2$, and
hyperbolic plane $H^2$, can be considered as three dif\/ferent
situations inside a family of Riemannian manifolds $M_{\k}^2 =
(S_\k^2, {\Bbb E}^2,H_\k^2)$ with the curvature $\k$ as a
parameter $\k\in \mathbb{R}$. In order to obtain mathematical
expressions valid for all the values of $\k$, it is convenient to
make use of the following $\kappa$-trigonometric functions
\begin{gather*}
\Cos_{\kappa}(x) = \left\{\begin{array}{lll} \cos{\sqrt{\kappa} x}
&{\rm if} &\kappa>0,
\\[1ex]
 1               &{\rm if} &\kappa=0,
\\[1ex]
\cosh\!{\sqrt{-\kappa} x}   &{\rm if} &\kappa<0, \end{array}
\right.{\qquad}
\Sin_{\kappa}(x) = \left\{\begin{array}{lll}
\dfrac{1}{\sqrt{\kappa}} \sin{\sqrt{\kappa} x}     &{\rm if}
&\kappa>0,
\\[1ex]
 x  &{\rm if} &\kappa=0,
\\
\dfrac{1}{\sqrt{-\kappa}}\sinh\!{\sqrt{-\kappa} x} &{\rm if}
&\kappa<0,
  \end{array}\right.
\end{gather*}
and the $\k$-dependent tangent function $\Tan_\k(x)$ def\/ined in
the natural way,  $\Tan_\k(x) = \Sin_{\k}(x)/\Cos_{\k}(x)$. The
fundamental properties of these curvature-dependent trigonometric
functions are
\begin{gather*}
   \Cos_{\k}^2(x)+ \k \Sin_{\k}^2(x) =1,
\end{gather*}
and
\begin{gather*}
   \Cos_{\k}(2x)=\Cos_{\k}^2(x) - \k \Sin_{\k}^2(x), \qquad
   \frac{d}{dx}\Sin_{\k}(x)=\Cos_{\k}(x),
   \\
   \Sin_{\k}(2x)=2\Sin_{\k}(x)\Cos_{\k}(x),  \qquad
   \frac{d}{dx}\Cos_{\k}(x)=-\k \Sin_{\k}(x).
\end{gather*}
If in the Lagrangian of the nonlinear oscillator in one dimension
\begin{gather*}
   L_\lambda(x,\dot x)  = \frac{1}{2}\ \frac{1}{1 + \lambda x^2}  \big(\dot{x}^2 -
\alpha^2 x^2\big),
\end{gather*}
we consider the change of variable $x=\Sin_\k(u)$, where
$\lambda=-\k$, we f\/ind that
\begin{gather*}1+\la  x^2=1-\k  x^2=1-\k  \Sin_\k^2(u)=\Cos_\k^2(u)
\end{gather*}
and as ${dx}/{du}=\Cos_\k(u)$, we see that
\begin{gather*}\dot x=\frac{dx}{du}  \dot u=\Cos_\k(u)  \dot u
\end{gather*}
and therefore  the Lagrangian becomes
\begin{gather*} L_\k(u,\dot u)  = \frac 12 \frac{\Cos_\k^2(u)  \dot u^2}{\Cos_\k^2(u)}-\frac 12
\frac{\alpha^2 \Sin_\k^2(u)}{\Cos_\k^2(u)}=\frac 12  \dot
u^2-\frac {\alpha^2}2 \Tan_\k^2(u),
\end{gather*}
the vector f\/ield $X_x$ turns out to be
\begin{gather*} X=\Cos_\k(u) \frac{du}{dx} \pd{}u=\pd {}u,
\end{gather*}
and the metric giving rise to the Lagrangian $g=du\otimes du$. We
note that the expression $\Tan_\k^2(u)$ of the above
one-dimensional potential appears as directly related with the
potential of the two-dimensional harmonic oscillator on a space
with curvature $\k$.

The expression of the dif\/ferential element of distance in
geodesic polar coordinates $(\rho,\phi)$ on the family
$M_{\k}^2=(S_\k^2,{\Bbb E}^2,H_\k^2)$, can be written as follows $
ds_{\k}^2 = d\rho^2 + \Sin_\k^2(\rho) d{\phi}^2$, so that it
reduces~to
\begin{gather*}
  ds_1^2 =    d\rho^2 + \big(\sin^2 \rho\big) d{\phi}^2,{\qquad}
  ds_0^2 =    d\rho^2 + \rho^2 d{\phi}^2,{\qquad}
  ds_{-1}^2 = d\rho^2 + \big(\sinh^2 \rho\big) d{\phi}^2,
\end{gather*}
in the three particular cases $\k=1,0,-1$ of the unit sphere,
Euclidean plane, and `unit' Lobachewski plane, respectively. Note
that $\rho$ denotes the distance along a geodesic on the manifold
$M_{\k}^2$; for example, in the spherical $\k>0$ case, $\rho$ is
the distance of the point to the origin (e.g., the North pole)
along a maximum circle.

Therefore, the Lagrangian for the geodesic (free) motion on the
spaces $(S_\k^2,{\Bbb E}^2,H_\k^2)$ is
\begin{gather*}
  L_\k(\rho,\phi,v_\rho,v_\phi) = T_\k(\rho,\phi,v_\rho,v_\phi) = \frac{1}{2} \big(v_\rho^2 + \Sin_\k^2(\rho)
  v_\phi^2\big),
\end{gather*}
and the Lagrangian for a general mechanical system (Riemannian
metric minus a potential) is
\begin{gather*}
  L_\k(\rho,\phi,v_\rho,v_\phi) = \frac{1}{2} \big( v_\rho^2 + \Sin_\k^2(\rho) v_{\phi}^2 \big)
  -  U(\rho,\phi,\k)  .
\end{gather*}
The spherical and hyperbolic harmonic oscillators are
characterized by the following Lagrangians with curvature $\k$
\cite{RaSa02I,RaSa03II}
\begin{gather*}
  L_\k(\rho,\phi,v_\rho,v_\phi) = \frac{1}{2} \left(v_\rho^2 + \Sin_\k^2(\rho) v_\phi^2\right)
  - \frac{1}{2} \om_0^2 \Tan_\k^2(\rho),
\end{gather*}
i.e. the harmonic oscillators on the unit sphere (Higgs
oscillator), on the Euclidean plane, or on the unit Lobachewski
plane, are:
\begin{gather*}
   U_1(\rho) = \frac{1}{2} \om_0^2 \tan^2\rho,{\qquad}
   U_0(\rho) = \frac{1}{2} \om_0^2 \rho^2,{\qquad}
   U_{-1}(\rho) = \frac{1}{2} \om_0^2 \tanh^2\rho .
\end{gather*}

Next we study the behavior of $L_\k$ under two dif\/ferent changes
of variables.


1.~If we consider the $\k$-dependent change $(\rho,\phi) \to
(r',\phi)$ given by $ r'  = \Tan_\k(\rho) $, then the Lagrangian
$L_\k$ becomes
\begin{gather*}
   L_{H\k} (r',\phi,v_{r'},v_\phi)= \frac{1}{2} \left(\frac{{v_{r'}}^{2}}{(1 + \k r'^2)^2} +
   \frac{{r'}^2 v_\phi^2}{(1 + \k r'^2)}  \right)
   - \frac{1}{2} \al^2 r'^2 .
\end{gather*}
This function coincides, in the spherical $\k>0$ case, with the
Lagrangian studied by Higgs in~\cite{Hi79}. In Cartesian
coordinates $(x,y)$ it reduces to
\begin{gather*}
   L_{H\k}(x,y,v_x,v_y)  = \frac{1}{2}  \frac{1}{(1 + \k  {r'}^2)}
   \bigl[ v_x^2 + v_y^2 + \k  (x v_y - y v_x)^2  \bigr]
   - \frac{1}{2} \al^2 r'^{2},\qquad
   r'^{2} = x^2+y^2.
\end{gather*}

2.~Let us consider the $\k$-dependent change $(\rho,\phi) \to
(r,\phi)$ given by $r  = \Sin_\k(\rho)$, $\la=- \k$. Then the
Lagrangian $L_\k$ becomes
\begin{gather*}
  L_\la(r,\phi,v_r,v_\phi) = \frac{1}{2} \left(\frac{v_r^2}{1 + \la r^2} +
  r^2v_\phi^2  \right)
  - \frac{\al^2}{2}\biggl(\frac{r^2}{1 + \la r^2} \biggr).
\end{gather*}
Therefore, if we change to Cartesian coordinates $(x,y)$ we arrive
to
\begin{gather*}
   L_\la(x,y,v_x,v_y)  = \frac{1}{2} \left(\frac{1}{1 + \la r^2} \right)
   \big[ v_x^2 + v_y^2 + \la (x v_y - y v_x)^2\big]
   - \frac{\al^2}{2}\left(\frac{r^2}{1 + \la r^2} \right),
   \\
   r^2 = x^2+y^2.
\end{gather*}
This function is just the Lagrangian obtained in~\cite{CaRaSS04} as the natural generalization of the
one-dimensional Lagrangian $L_\la(x,v_x)$ for the nonlinear
equation (\ref{nloeq}) of Mathews and Lakshmanan.

We thus have three dif\/ferent and alternative ways of describing
the harmonic oscillator on spaces of constant curvature: the
original $\k$-dependent trigonometric (hyperbolic) Lagrangian~$L_\k$ and the two other approaches, $L_{H\k}$ and $L_\la$,
obtained from it.

   The Higgs approach \cite{Hi79} has been studied by many authors (see e.g.~\cite{BoDaK93,BoDaK94} and refe\-rences therein) mainly
in relation with the theory of dynamical symmetries. Concerning
the $\la$-dependent Lagrangian $L_\la$, it has similarities with
$L_{H\k}$ but it does not coincide with it. In the model of Higgs
$\k$ (or $\la$) is present in the kinetic term $T$ in a
dif\/ferent way and the potential $V$ appears as $\k$-independent;
this af\/fects to the Hamiltonian formalism. On the other side
each one of these three formalisms can be used for
 study of the $\k$-dependent version of the S-W system.
In the language of $L_\k$ the potential, that was studied in
\cite{RaSa99,RaSa03II}, is given by
\begin{gather*}
U_a(r,\phi,\k) = k_1 U_a^1 + k_2 U_a^2 + k_3 U_a^3 + k_0,\qquad
k_1=\frac{1}{2} \om_0^2,
\\
U_a^1 = \Tan_\k^2(\rho),\qquad U_a^2 =
\frac{1}{(\Sin_\k(\rho)\cos{\phi})^2},\qquad U_a^3 =
\frac{1}{(\Sin_\k(\rho)\sin{\phi})^2},
\end{gather*}
with integrals of mot\-ion given by
\begin{gather*}
   I_1(\k) = P_1^2(\k)
            + {\om_0^2} \bigl(\Tan_\k(\rho) \cos{\phi}\bigr)^2
            + \frac{2 k_2}{\bigl(\Tan_\k(\rho) \cos{\phi}\bigr)^2},
            \\
   I_2(\k) = P_2^2(\k)
            + {\om_0^2} \bigl(\Tan_\k(\rho) \sin{\phi}\bigr)^2
            + \frac{2 k_3}{\bigl(\Tan_\k(\rho) \sin{\phi}\bigr)^2},
            \\
   I_3(\k) = J^2(\k) + \frac{2 k_2}{\cos^2{\phi}}
            + \frac{2 k_3}{\sin^2{\phi}},
\end{gather*}
with $P_1(\k)$, $P_2(\k)$ and $J(\k)$ given by
\begin{gather*}
   P_1(\k) = (\cos{\phi}) v_\rho - (\Cos_\k(\rho)\Sin_\k(\rho)\sin{\phi}) v_{\phi},
\\
   P_2(\k)=  (\sin{\phi}) v_\rho + (\Cos_\k(\rho)\Sin_\k(\rho)\cos{\phi}) v_{\phi},
\qquad
   J(\k) =\Sin_{\k}^2(\rho) v_{\phi}.
{\nonumber}
\end{gather*}
This S-W system has been studied, using a dif\/ferent approach, in
the two- and three- dimensional sphere in \cite{GrPoSi95}, in the
two-dimensional hyperboloid in \cite{KaMiPo97} and in the complex
two-sphere in  \cite{KaKrPo01}.

One of the advantages of the $L_\la$ approach is that the
Euler--Lagrange equations can be directly solved and the general
solution has a rather simple form that can be interpreted as
``quasi-harmonic" nonlinear oscillations  \cite{CaRaSS04}; other
important advantage is that it is very appropriate for the study
of the quantum oscillator.

In what follows we will focus our attention on the quantum
Hamiltonian dynamics determined by the $\la$-dependent Lagrangian
$L_\la$.

\section{The one-dimensional  quantum nonlinear oscillator}

Let us consider the quantum case for $n=1$. The problem is to
def\/ine the quantum operator def\/ining the Hamiltonian of this
position-dependent mass system, because the mass and the momentum
$P$ do not commute and this fact gives rise to an ambiguity in the
ordering of factors.

Instead of using traditional procedures as Weyl ordering we shall
develop an alternative method of quantization \cite{CRS04}. We
f\/irst remark that the vector f\/ield
\begin{gather*}
   X_x(\lambda) = \sqrt{ 1+\lambda x^2 }  \pd{}{x},
\end{gather*}
which was a Killing vector for the metric corresponding to the
kinetic energy
\begin{gather*}g=\big(1 + \la x^2\big)^{-1}  dx\otimes dx
\end{gather*}
and generates the translations in this Riemann space, does not
leave invariant the natural measure in the real line but the only
invariant measures are the multiples of
\begin{gather*}d\mu=\big(1 + \la x^2\big)^{-1/2}  dx.
\end{gather*}
This suggests us to consider the Hilbert space
$\mathcal{L}^2(\mathbb{R},d\mu)$ and the remarkable fact is that
the adjoint of the dif\/ferential operator $\sqrt{ 1+\la x^2 }
 \partial/\partial{x}$ in such space is the opposite of such
operator.

The Legendre transformation corresponding to this deformed kinetic
energy is def\/ined by
\begin{gather*}p=\frac {v_x}{1 + \la x^2}
\end{gather*}
and then the Hamiltonian function for the free particle is given
by
\begin{gather*}H=\big(1 + \lambda x^2\big)  \frac{p^2}2
   =\frac 12 \left(\sqrt{1 + \lambda x^2} p \right)^2.
\end{gather*}

The usual prescription of canonical quantization does not present
any ambiguity because the linear operator (we put $\hbar=1$)
\begin{gather*}\wh P=-i   \sqrt{1 + \la x^2} \pd{}x
\end{gather*}
is self-adjoint in the space  $\mathcal{L}^2({\mathbb{R}},d\mu)$.
Note that with the above mentioned change of coordinates,
$x=\Sin_\k(u)$, $d\mu=du$ and{\samepage
\begin{gather*}\pd{}u=\Cos_\k(u) \pd{}x
   = \sqrt{1-\k \Sin_\k^2(u)} \pd{}x = \sqrt{1 + \la x^2} \pd{}x
\end{gather*}
is self-adjoint in the space  $\mathcal{L}^2({\mathbb{R}},du)$,
and then $\wh P$ generates translations in these coordinates $u$.}

The quantum Hamiltonian operator of the free particle is
\begin{gather*}\widehat H
   = -\frac 12 \left( \sqrt{1 + \lambda x^2} \pd{}x \right)^2
   = -\frac 12 \big(1 + \lambda x^2\big) \pd{^2}{x^2}-\frac 12\lambda x \pd{} x
\end{gather*}
and in presence of an interaction  $V_1(x)$ the Hamiltonian will
be
\begin{gather}
   \widehat{H}_1 = -\frac 12 \big(1 + \la x^2\big) \frac{d^2}{dx^2}
       -\frac 12 \la x \frac{d}{dx} +V_1(x)  .
\end{gather}
We are interested in the case of the nonlinear oscillator for
which
\begin{gather}
\widehat{H}_1 = -\frac 12 \big(1 + \lambda x^2\big)
\frac{d^2}{dx^2}
       -\frac 12 \lambda x \frac{d}{dx} +
\frac{1}{2}  \frac{\alpha^2 x^2}{1 + \lambda x^2}  .
\end{gather}
If $m$ and $\hbar $ are taken into account, and with the change of
parameter, to be more clear later, given by
\begin{gather*}
   \alpha^2=\beta\left(\beta+\frac\hbar m \lambda\right)
\end{gather*}
it is enough to def\/ine the dimensionless variables
\begin{gather*}
y=\sqrt{\frac{m\beta}\hbar},\qquad
   \Lambda=\frac\hbar{m \beta} \lambda
\end{gather*}
and then $\widehat{H}_1$ is written as
\begin{gather*}
\widehat{H}_1   = \beta \hbar \left[ -\frac 12 \big(1 + \Lambda
y^2\big) \frac{d^2}{dy^2}-\frac 12 \Lambda y \frac{d}{dy}+\frac 12
(1+\Lambda) \frac{y^2}{1 + \lambda y^2}\right].
\end{gather*}
The time-independent Schr\"odinger equation
\begin{gather*}
\widehat{H}_1 \Psi=E \Psi,\qquad  E=\hbar\beta {\cal E}
\end{gather*}
becomes
\begin{gather*}
   \left[ -\frac 12 \big(1 + \Lambda y^2\big)
    \frac{d^2}{dy^2}-\frac 12 \Lambda y \frac{d}{dy}
   +\frac 12  (1+\Lambda) \frac{y^2}{1 + \lambda y^2}\right]\Psi
   = {\cal E} \Psi.
\end{gather*}

With the change of variable \cite{CRS07a}
\begin{gather*}
\Psi(y,\Lambda)=\varphi(y,\Lambda)  \big(1+\Lambda
y^2\big)^{-1/(2\Lambda)}
\end{gather*}
the eigenvalue equation becomes
\begin{gather*}
\big(1+\Lambda  y^2\big)\varphi''   + (\Lambda-2)  y \varphi'+(2
{\cal E}-1)\varphi = 0
\end{gather*}
and assuming a power series development
\begin{gather*}
\varphi(y)=\sum_{n\geq 0}a_n  y^n
\end{gather*}
   the following recursion relation for the coef\/f\/icients is obtained:
\begin{gather*}
a_{n+2}=(-1)^n\frac {a_n}{(n+2)(n+1)}   [n(\Lambda n-2+(2 {\cal
E}-1)].
\end{gather*}
Therefore the general solution is determined by the values of the
coef\/f\/icients $a_0$ and $a_1$. In particular we can write the
solutions $y_0$ and $y_1$ determined by $a_0=1,  a_1=0$ and
$a_0=1,  a_1=0$, respectively. The solution $y_0$ is an even
function while the $y_1$ function is odd. The general solution
will be written as $y=a_0  y_0+a_1  y_1$.

The convergence radius of such power series is
\begin{gather*}
R=\frac 1{\sqrt {|\Lambda|}} \qquad\mbox{because}\quad
\lim_{n\to\infty}\left|\frac{a_{n+2}}{a_n}\right|=|\Lambda|.
\end{gather*}
This series  $y=a_0  y_0+a_1  y_1$ reduces to a polynomial of
degree $p$  when one of the two coef\/f\/icients vanish and that
means that there exists a positive integer number  $p$ such that
$2 {\cal E}-1=2  p-\Lambda  p^2$, and then
\begin{gather*}{\cal E}_p=p\left(1-\Lambda  \frac p2\right)+\frac 12.
\end{gather*}
The polynomial solutions can be easily found and they have a form
quite similar to the corresponding Hermite polynomials.

\section{Factorization method and shape-invariance}

   The spectrum of the harmonic oscillator can be found using
the factorization method we shall describe in this section
\cite{CRS04}. Actually the Hamiltonian can be factorized and we
arrive to a shape-invariant Hamiltonian for which the full
point-spectrum can be found by algebraic methods. We take $\hbar=
\omega=1$.  The eigenvalue problem is  (up to  a factor $1/2$)
\begin{gather*}
H\psi_n =\left(-\frac {d^2}{dx^2}+x^2\right)\psi_n=(2n+1)\psi _n,
\end{gather*}
where the Hamiltonian $H$ is such that $H-1=a^{\dag} a$, with
$a=\left( d/{dx}+x\right)$, $a^{\dag}=\left(- d/{dx}+x\right)$,
and then the ground state is found from
\begin{gather*}
a\psi_0=\left(\frac d{dx}+x\right)\psi_0=0,
\end{gather*}
i.e. $\psi_0\propto e^{-{x^2}/2}$.  The other eigenstates can be
found by applying an iterative way  the creation operator
$a^{\dag}$ to the ground state:
\begin{gather*}
\psi_n=\frac 1{\sqrt{2^nn!}} a^{\dag n}\psi_0\propto    H_n(x)
e^{-x^2/2},
\end{gather*}
with $H_n(x)$ being the Hermite polynomials.

A similar procedure can be used for the quantum nonlinear
oscillator case.  We should look for a function $W(x)$, to be
called super-potential, in such  a way that the operator $A$ and
its adjoint operator $A^+$, given by
\begin{gather*}
A  =  \dfrac 1{\sqrt 2}\left(\sqrt{1 + \la x^2}\dfrac{d}{dx}
+W(x)\right), \qquad A^+ =\dfrac 1{\sqrt 2}\left( -\sqrt{ 1 + \la
x^2 }\dfrac{d}{dx} +W(x)\right),
\end{gather*}
are such that $\widehat H_1=A^+  A$, i.e.
\begin{gather*}
\widehat H_1 = A^+A =\frac 12   \left[-\sqrt{1 +  \la x^2}
\frac{d}{dx} +W(x)\right] \left[\sqrt{1 + \la x^2} \frac{d}{dx}
+W(x) \right].
\end{gather*}
Therefore, in order to the Hamiltonian so found be that of the
deformed quadratic energy term together with a potential $V_1$,
the  super-potential function $W$ must satisfy
\begin{gather*}
\sqrt{1 + \la x^2}W' - W^2 +2  V_1 = 0.
\end{gather*}
We can def\/ine a new quantum Hamiltonian operator
\begin{gather*}
\widehat H_2 = AA^+ =\left[\sqrt{1 + \la x^2} \frac{d}{dx} +W(x)
\right] \left[-\sqrt{1 + \la x^2} \frac{d}{dx} +W(x) \right]
\end{gather*}
which is called the partner Hamiltonian. The new potential $V_2$
is given in terms of  $W$ by
\begin{gather*}
   V_2 = \frac 12\left(\sqrt{ 1 + \la x^2 } W' + g W^2\right).
\end{gather*}

The important fact is that
\begin{gather*}A  \widehat H_1 = \widehat H_2  A,\qquad
   A^+  \widehat H_2= \widehat H_1 A^+ .
\end{gather*}
When $\widehat H_1\ket\Psi = E \ket\Psi$,  then,  $\widehat H_2 A
\ket\Psi = A  \widehat H_1 \ket\Psi=E A \ket\Psi$.  If
$A\ket\Psi\ne 0$, $A\ket\Psi$ is an eigenvector of  $\widehat H_2$
corresponding to the same eigenvalue $E$, and similarly, if
$\ket\Phi$ is an eigenvector of $\widehat H_2$ with eigenvalue $E$
and such that $A^+ \ket\Phi\ne 0$, then $A^+ \ket\Phi$ is an
eigenvector of $\widehat H_1$ corresponding to the same eigenvalue
$E$.

The spectra of $\widehat H_1$ and $\widehat H_2$ are then almost
identical, the only dif\/ferences are when either $\ket\Psi$ is an
eigenvector of $\widehat H_1$ but  $A \ket\Psi\ne 0$, or
$\ket\Phi$  is an eigenvector of  $\widehat H_2$ for which $A^+
\ket\Phi= 0$.

Some parameters may appear in the expression of $V_1$, and the
super-potential function $W$ will also depend on them. The most
important case is when the explicit forms of the potential and its
partner are quite similar and only dif\/fer in the values of the
parameters. In this case  we say that the problem has shape
invariance (see e.g.~\cite{CaRa00Rmp}).

Suppose that a quantum Hamiltonian $\widehat H_1(\alpha)$ admits a
factorization $\widehat H_1(\alpha) = A^+(\alpha) A(\alpha)$ in
such a way that the partner Hamiltonian $\widehat H_2(\alpha)$ is
of the same form as $\widehat H_1(\alpha)$ but for a dif\/ferent
value of the parameter $\alpha$.  More specif\/ically, there
exists a function $f$ such that
\begin{gather*}\widehat H_2(\alpha)=\widehat H_1(\alpha_1)+R(\alpha_1),
\end{gather*}
where  $\alpha_1=f(\alpha)$ and  $R(\alpha)$ is a constant
depending on the parameter $\alpha$.  In this case
Gendenshte\"{\i}n developed  a method for exact computing of all
the spectrum of $\widehat H_1$ \cite{{CaRa00Rmp},{Ge83},{GeKr85}}.
First, the bound state $\ket{\Psi_0}$ is found by solving
$A(\alpha)\ket{\Psi_0(\alpha)} = 0$, and has a zero energy.  Then,
$\ket{\Psi_0(\alpha_1)}$ is an eigenstate of $\widehat
H_2(\alpha)$ with $E_1=R(\alpha_1)$, because
\begin{gather*}
   \widehat H_2(\alpha)\ket{\Psi_0(\alpha_1)}
   =(\widehat H_1(\alpha_1)+R(\alpha_1)) \ket{\Psi_0(\alpha_1)}
   = R(\alpha_1)\ket{\Psi_0(\alpha_1)},
\end{gather*}
and  $A^{\dag}(\alpha)\ket{\Psi_0(\alpha_1)}$ is the f\/irst
excited state of  $\widehat H_1(\alpha)$, with   energy
$E_1=R(\alpha_1)$, because:
\begin{gather}
   \widehat H_1(\alpha)A^{\dag}(\alpha)\ket{\Psi_0(\alpha_1)}=
   A^{\dag}(\alpha)(\widehat
   H_1(\alpha_1)+R(\alpha_1))\ket{\Psi_0(\alpha_1)}
   = R(\alpha_1)A^{\dag}(\alpha)\ket{\Psi_0(\alpha_1)}.\nonumber
\end{gather}
Iterating the process   we  f\/ind the sequence of energies for
   $\widehat H_1(\alpha) $
\begin{gather*}
   E_k = \sum_{j=1}^k R(\alpha_j),\qquad E_0=0,
\end{gather*}
the corresponding eigenfunctions being{\samepage
\begin{gather*}
   \ket{\Psi_n(x,\alpha_0)}=A^{\dag}(\alpha_0)A^{\dag}(\alpha_1)\cdots
   A^{\dag}(\alpha_{n-1})\ket{\Psi_0(x,\alpha_n)},
\end{gather*}
where $\alpha_0=\alpha$ and  $\alpha_{j+1}=f(\alpha_j)$, namely,
$\alpha_k=f^k(\alpha_0)=f^k(\alpha)$.}

Coming back to the nonlinear oscillator case, if $\beta\in
{\mathbb{R}}$,  we def\/ine the linear operator in
$\mathcal{L}^2({\mathbb{R}},d\mu)$
\begin{gather*}
   A  =  \dfrac 1{\sqrt 2}\left(\sqrt{1 + \lambda x^2}
   \dfrac{d}{dx} +\dfrac{\beta x}{\sqrt{1 + \lambda x^2}} \right),
\end{gather*}
for which its adjoint operator is
\begin{gather*}
   A^+ =\dfrac 1{\sqrt 2}\left( -\sqrt{1 + \lambda x^2}\dfrac{d}{dx}
   + \dfrac{\beta x}{\sqrt{1 + \lambda x^2}} \right)   .
\end{gather*}
Then, we f\/ind that if $\widehat H'_1=\widehat H_1-(1/2)\beta$,
\begin{gather*}
\widehat H'_1 = A^+A = -\frac{1}{2}\big(1 + \lambda
x^2\big)\frac{d^2}{dx^2}- \frac{1}{2}\lambda x\frac{d}{dx} +
\frac{1}{2}\beta(\beta+\lambda)\frac{x^2}{1 + \lambda x^2}-
\frac{1}{2} \beta,
\\
\widehat H'_2=AA^+ = - \frac{1}{2}\big(1 + \lambda
x^2\big)\frac{d^2}{dx^2} - \frac{1}{2}\lambda x \frac{d}{dx} +
\frac{1}{2}\beta(\beta-\lambda)\frac{x^2}{1 + \lambda x^2} +
\frac{1}{2} \beta.
\end{gather*}

This shows that there is one positive number $\beta$ for which the
Hamiltonian $\widehat H'_1$ of the quantum non-linear oscillator
admits a factorization with the parameters $\alpha$ and $\beta $
related by $\alpha^2=\beta(\beta+\lambda)$.

In our preceding case, the parameter being $\beta$, when comparing
$\widehat H'_1$  with its partner, as
\begin{gather*}
\widehat  H'_1(\beta-\lambda) = -\frac{1}{2}\left[\big(1 + \la
x^2\big)\frac{d^2}{dx^2} +\la x\frac{d}{dx} \right]
+\frac{1}{2}(\beta-\lambda)\beta\left(\frac{x^2}{1 +\lambda
x^2}\right)-\frac{1}{2}(\beta-\lambda),
\end{gather*}
then we see that
\begin{gather*}
\widehat  H'_1(\beta-\lambda) = \left(\widehat H'_2(\beta) -
\frac{1}{2}\beta\right) -\frac{1}{2}(\beta-\lambda)
\end{gather*}
and therefore
\begin{gather*}
\widehat H'_2(\beta)  = \widehat  H'_1(f(\beta))  + \beta -
\frac{1}{2} \la
\end{gather*}
where $f$ is the function $f(\beta) = \beta-\lambda$. If $R$ is
def\/ined by $R(\beta) = \beta + (1/2)$, then
\begin{gather*}
\widehat  H'_2(\beta) = \widehat H'_1(\beta_1) + R(\beta_1).
\end{gather*}
This shows that, as the quantum non-linear oscillator is shape
invariant, we can develop the~me\-thod sketched before: First, the
eigenvector $\ket{\Psi_0}$ is determined by the condition
$A(\beta_0) \ket{\Psi_0} = 0$. More specif\/ically, we should
solve the dif\/ferential equation
\begin{gather*}
\frac{d}{dx} \Psi_0 +\beta \frac{x}{1 +\la x^2} \Psi_0 =  0
\end{gather*}
and therefore the wave function of the fundamental state must be
proportional to
\begin{gather*}
\Psi_0 =  \frac{1}{( 1 + \lambda x^2)^{ r_0}},\qquad r_0 = \frac
\beta{2\lambda}.
\end{gather*}
The energies of the f\/irst  excited states will be
\begin{gather*}
E'_1=R(\beta_1)=\beta-\lambda+\frac\lambda2
\end{gather*}
and iterating the process we get
\begin{gather*}
E'_n=\sum_{k=1}^nR(\beta_k)=\sum_{k=1}^n\left(\beta_k+\frac\lambda
2\right) =\sum_{k=1}^n\left(\beta-\lambda k+\frac\lambda 2\right),
\end{gather*}
and therefore,
\begin{gather*}
E'_n=n \beta+\lambda\left[\frac n2-\sum_{k=1}^n k\right] = n
\beta-\frac{n^2}2 \lambda.
\end{gather*}
The energy of the eigenstates of $\widehat H_1=\widehat
H'_1+(1/2)\beta$ will be given by
\begin{gather*}
E_n=n \beta-\frac {n^2}2 \lambda+\frac 12 \beta.
\end{gather*}
The method  also provides us the corresponding eigenfunctions as
\begin{gather*}
\ket{\Psi_1(\beta)}=A^+(\beta)\ket{\Psi_0(\beta_1)},
\\
\ldots\ldots\ldots\ldots\ldots\ldots\ldots\ldots\ldots,
\\
\ket{ \Psi_n(\beta)}=A^+(\beta)A^+(\beta_1)\cdots
A^+(\beta_{n-1})\ket{\Psi_0(\beta_n)}.
\end{gather*}
There is a clear dif\/ference between the case $\lambda>0$ and the
case $\lambda<0$. Note that the lowest value for $E'_n$ is
$E'_0=0$.  Therefore: If $\lambda>0$ only values such that
\begin{gather*}
\beta-\lambda  \frac n2\geq 0\ \Longrightarrow \  n\leq\frac
{2\beta}\lambda
\end{gather*}
are allowed. The eigenvalues are not equally spaced.  On the
contrary, when  $\lambda<0$ all natural numbers are allowed
for~$n$

\section{Quantization of the 2-dimensional nonlinear\\ harmonic oscillator}

There exist relatively few examples of quantum Hamiltonians in two
dimensions whose spectrum can be fully determined by algebraic
methods.  We shall show that the deformed nonlinear oscillator is
one of such examples. Furthermore, this can be done in dif\/ferent
ways \cite{CRS07b}.

First, one can check that all the  measures   $d\mu=\rho(x,y)
dx\wedge dy$  invariant under the vector f\/ields
\begin{gather*}X_1=\sqrt{1+\lambda  r^2}  \pd {}{x},\qquad X_2=\sqrt{1+\lambda  r^2}
\pd {}{y},
\end{gather*}
are  proportional to
\begin{gather*}d\mu=\frac 1{\sqrt{1+\lambda  r^2}} dx\wedge dy.
\end{gather*}

Therefore we shall consider the Hilbert space
${\mathcal{L}}^2({\mathbb{R}}^2,d\mu)$.

The form of Killing vectors $X_1$ and $X_2$ generating
`translations' suggests us to take the momenta operators
\begin{gather*}\wh P_x=-i  \hbar \sqrt{1+\lambda  r^2} \pd{}x,\qquad
\wh P_y=-i  \hbar \sqrt{1+\lambda  r^2} \pd{}y,
\end{gather*}
and then the quantum Hamiltonian is
\begin{gather*}
\wh H=-\frac{\hbar^2}{2m}\left(\big(1+\lambda
r^2\big)\pd{^2}{x^2}+\lambda x\pd{}x\right)
-\frac{\hbar^2}{2m}\left(\big(1+\lambda
r^2\big)\pd{^2}{y^2}+\lambda y\pd{}y\right)
\\ \phantom{\wh H=}
{}+\lambda \frac{\hbar^2}{2m}\left(x^2 \pd{^2}{y^2} +y^2
\pd{^2}{x^2}-2 x y\pd{^2}{x \partial y}-x\pd{}x-y\pd{}y
\right)+\frac 12   g  \frac{r^2}{1+\lambda  r^2},
\end{gather*}
which can be written as
\begin{gather*}\wh H=\wh H_1+\wh H_2-\lambda\wh J^2,
\end{gather*}
with
\begin{gather*}
\wh H_1=-\frac{\hbar^2}{2m}\left(\big(1+\lambda
r^2\big)\pd{^2}{x^2}+\lambda x\pd{}x\right)+\frac 12   g
\frac{x^2}{1+\lambda  r^2},
\\
\wh H_2=-\frac{\hbar^2}{2m}\left(\big(1+\lambda
r^2\big)\pd{^2}{y^2}+\lambda y\pd{}y\right)+\frac 12   g
\frac{y^2}{1+\lambda  r^2},
    \\
\wh J^2=-\frac{\hbar^2}{2m}\left(x^2 \pd{^2}{y^2}+y^2
\pd{^2}{x^2}-2 x y\pd{^2}{x \partial
y}-x\pd{}x-y\pd{}y\right)+\frac 12   g  \frac{r^2}{1+\lambda
r^2}.
\end{gather*}
Note that  each term commutes with the sum of the other two and
therefore with $\wh H$. Consequently,  we can consider three
dif\/ferent (complete) systems of compatible observables:
\begin{gather*}
\{\wh H_1,\wh H_2-\lambda  J^2\},\qquad \{\wh H_1-\lambda  J^2,\wh
H_2\},\qquad \{\wh H_1+\wh H_2, J\}.
\end{gather*}

Therefore we should solve one of these spectral problems:
\begin{gather*}
{\rm A)} \ \ \wh H_1  \Psi(E_1,E_{2j})=E_1  \Psi(E_1,E_{2j}),\quad
(\wh H_2-\lambda  J^2) \Psi(E_1,E_{2j})=E_{2j}\Psi(E_1,E_{2j});\\
{\rm B)} \ \ (\wh H_1-\lambda  J^2) \Psi(E_{1j},E_{2})=E_{1j}
\Psi(E_{1j},E_{2}),\quad
\wh H_2  \Psi(E_{1j},E_{2})=E_{2}  \Psi(E_{1j},E_{2});\\
{\rm C)} \ \ \wh H_1 \Psi(E_1,E_{2j})=E_1  \Psi(E_1,E_{2j}),\quad
(\wh H_2-\lambda  J^2) \Psi(E_1,E_{2j})=E_{2j}  \Psi(E_1,E_{2j}).
\end{gather*}

As in the one-dimensional case, it is convenient to use $g = m
\alpha^2 +\lambda \hbar \alpha$ and dimensionless variables
\begin{gather*}
x=\sqrt{\frac\hbar{m \alpha}}\ \tilde x,\qquad
y=\sqrt{\frac\hbar{m \alpha}}\ \tilde y, \qquad \lambda =\frac {m
\alpha}\hbar\ \Lambda,\qquad E=\hbar \alpha   e
\end{gather*}
for which $1+\lambda  r^2=1+\Lambda \tilde r^2$ and the
Schr\"odinger equation becomes
\begin{gather*}
-\left(\big(1+\Lambda  r^2\big)\pd{^2}{x^2}+\Lambda
x\pd{}x\right)\Psi -\frac{\hbar^2}{2m}\left(\big(1+\Lambda
r^2\big)\pd{^2}{y^2}+\Lambda y\pd{}y\right)\Psi
\\ \qquad
{}+\Lambda \frac{\hbar^2}{2m}\left(x^2 \pd{^2}{y^2}+y^2
\pd{^2}{x^2}-2 x y\pd{^2}{x \partial
y}-x\pd{}x-y\pd{}y\right)\Psi+\frac 12
(1+\Lambda)\frac{r^2}{1+\Lambda r^2}\Psi=e \Psi.
\end{gather*}

A)\ As the Hamilton--Jacobi equation separates in coordinates
$(z_x,y)$ we shall use such coordinates to write the Schr\"odinger
equation ($z$ is used instead of $z_x$)
\begin{gather*}
-\frac 12\left(\frac{1+\Lambda z^2}{1+\Lambda
y^2}\pd{^2}{z^2}+\frac{\Lambda z}{1+\Lambda
y^2}\pd{}{z}\right)\Psi
\\ \qquad
{}-\frac 12\left((1+\Lambda y^2)\pd{^2}{y ^2}+2\Lambda
y\pd{}{z}\biggr) \Psi +\frac 12\frac {1+\Lambda}{1+\Lambda
y^2}\biggl(\frac{z^2}{1+\Lambda  z^2}+y^2\right) \Psi =e\Psi,
\end{gather*}
and assuming a factorization for $\Psi(z,y)$  of the form
$\Psi(z,y)=Z(z)Y(y)$ we f\/ind the following equations:
\begin{gather*}
-\frac 12 \left(\big(1+\Lambda z^2\big)Z''+\Lambda z
Z'\right)+\frac 12 (1+\Lambda)\frac{z^2}{1+\Lambda  z^2} Z=\mu  Z,
\\
\qquad{} -\frac 12 \left(\big(1+\Lambda y^2\big)^2Y''+2 \Lambda y
\big(1+\Lambda y^2\big)Y'\right) +\left(\frac12 (1+\Lambda)
y^2-\big(1+\Lambda y^2\big) e\right) Y=-\mu Y.
\end{gather*}

In this way the two-dimensional problem has been decoupled in two
1-dimensional equations. The f\/irst one is the corresponding one
to a 1-dimensional system. The second one however is a bit
dif\/ferent because includes the contribution of the angular
momentum.

If  a new parameter $\nu=e-\mu$ is introduced the second equation
becomes
\begin{gather*}\big(1+\Lambda
y^2\big)Y''+2 \Lambda y Y'-(1+\Lambda-2 \Lambda
\mu)\frac{y^2}{1+\Lambda
    y^2}Y+2 \nu Y=0,
\end{gather*}
i.e.\ def\/ining $G_\mu^2=1+(1-2\mu)\Lambda$,
\begin{gather*}\big(1+\Lambda y^2\big)Y''+2 \Lambda y Y'-G_\mu^2 \frac{y^2}{1+\Lambda
    y^2}Y+2 \nu Y=0,
\end{gather*}
and then writing
\begin{gather*}Y(\Lambda,\mu)=q(y,\Lambda) \big(1+\Lambda  y^2\big)^{-G_\mu/(2\Lambda)},
\end{gather*}
it becomes
\begin{gather*}\big(1+\Lambda y^2\big) q''+2(\lambda-G_\mu) yq'+(2 \nu-G_\mu)q=0
\end{gather*}
which is a deformation of the Hermite equation.

Assuming the power expansion
\begin{gather*}
q(y,\Lambda)=\sum_{n=0}^\infty c_n(\Lambda)  y^n
\end{gather*}
we obtain the recursion relation
\begin{gather*}c_{n+2}=-\frac{c_n}{(n+2)(n+1)}
   \left(\Lambda n(n-1) -G_\mu(2n+1)+2  \nu\right).
\end{gather*}
Therefore the general solution is determined by the values of the
coef\/f\/icients $c_0$ and $c_1$. In particular we can write the
solutions $q_0$ and $q_1$ determined by $c_0=1$,  $c_1=0$ and
$c_0=1$,  $c_1=0$, respectively. The solution $q_0$ is an even
function while the $q_1$ function is odd. The general solution is
$q=c_0  q_0+c_1  q_1$.

The radius of convergence of both power series def\/ining $q_0$
and $q_1$ is
\begin{gather*}R=\frac 1{\sqrt{|\Lambda|}}.
\end{gather*}

The solution reduces to a polynomial of degree $n$ if one of the
coef\/f\/icients is zero and there exists an integer number $n$
related with  $\nu$ as follows:
\begin{gather*}2 \nu=G_\mu(2n+1)- n(n+1)\Lambda .
\end{gather*}

One can study the properties of these polynomial solutions which
have  a lot of similarities with the Hermite polynomials. The
associated Sturm--Liouville problem plays a relevant r\^ole.

In summary, the bound states for this system have energies
\begin{gather*}e_{m,n} = \mu_m+\nu_n=(m+n+1)\left(1-\frac12(m+n)\lambda\right)
\end{gather*}
and wave functions given by
\begin{gather*}\Psi_{m,n}(z,y) =  Z_m(z)  Y_n(y),\qquad
z=\frac x{\sqrt{1+\Lambda y^2}}  .
\end{gather*}

We f\/inally mention that the properties of these $\la$-dependent
Hermite polynomials are discussed in \cite{CRS07a,CRS07b}. It is
proved the orthogonality as well as the existence of a
$\la$-dependent Rodrigues formula, a generating function and
$\la$-dependent recursion relations between polynomials of
dif\/ferent orders.

\subsection*{Acknowledgments}
Partial f\/inancial support of research  projects BFM-2003-02532,
FPA-2003-02948, MTM-2005-09183, DGA E24/1 and VA013C05 is
acknowledged.

\pdfbookmark[1]{References}{ref}
\LastPageEnding

\end{document}